\documentclass[aps, pra, twocolumn, showpacs, nofootinbib, superscriptaddress]{revtex4-1}
\usepackage{graphicx}
\usepackage{amsmath}
\usepackage{amsfonts}
\usepackage{amssymb}
\usepackage{hyperref}

\newcommand{\ket}[1]{|{#1}\rangle}

\newcommand{\beq}{\begin{equation}}
\newcommand{\eeq}{\end{equation}}
\newcommand{\grad}{\nabla}

\begin{document}

\title{Ion chains in high-finesse cavities}

\author{Cecilia Cormick} 
\affiliation{Theoretische Physik, Universit\"at des Saarlandes, D-66123 Saarbr\"ucken, Germany} 

\author{Giovanna Morigi} 
\affiliation{Theoretische Physik, Universit\"at des Saarlandes, D-66123 Saarbr\"ucken, Germany} 

\date{\today}

\begin{abstract}
We analyze the dynamics of a chain of singly-charged ions confined in a linear Paul trap and which couple with the mode
of a high-finesse optical resonator. In these settings the ions interact via the Coulomb repulsion and are subject to
the mechanical forces due to scattering of cavity photons. We show that the interplay of these interactions can give
rise to bistable equilibrium configurations, into which the chain can be cooled by cavity-enhanced photon scattering. We
characterize the resulting equilibrium structures by determining the stationary state in the semiclassical limit for
both cavity field and crystal motion. The mean occupation of the vibrational modes at steady state is evaluated, showing
that the vibrational modes coupled to the cavity can be simultaneously cooled to low occupation numbers. It is also
found that at steady state the vibrations are entangled with the cavity field fluctuations. The entanglement is
quantified by means of the logarithmic negativity. The spectrum of the light at 
the cavity output is evaluated and the features signaling entanglement are identified. 
\end{abstract}


\maketitle

\section{Introduction}

High-finesse resonators are fundamental elements in quantum-optical setups: The strong coupling between single photons
and single atoms allows one to achieve remarkable levels of control \cite{Haroche_RMP_2001, Haroche_book, Walther_2006,
Kimble,Rempe} which offer promising perspectives for quantum technological applications \cite{Ritter_2012}. In presence
of many atoms, multiple scattering of cavity photons induces an effective atom-atom interaction, which in a single-mode
cavity is infinitely ranged \cite{Domokos_Ritsch_J0SAB_2003} and can cool the atoms into self-organized patterns
\cite{CARL_Theory, CARL_Exp, Domokos_Ritsch_2002, Chan:03, Black:03, Hemmerich:Science2012}. When the scattering
processes are prevailingly coherent, the coupling with the cavity field gives rise to a conservative periodic potential
in which ultracold atoms can spatially order \cite{Domokos_EPJD,Esslinger}. In this setting it has been shown that a
quantum gas of bosonic atoms can exhibit supersolidity \cite{Esslinger}. 

Most experiments realized so far employed gases of neutral atoms. The interaction is there essentially due to the
cavity-mediated potential, while collisions at ultralow temperatures can be described by a contact interaction
\cite{Stringari_RMP}. The scenario is quite different when the particles coupling with the resonator are singly-charged
ions \cite{Herskind_NatPhys_2009, Albert_PRA_2012}. In this case, the long-range Coulomb repulsion is dominant, while
the mechanical forces associated with multiple scattering of a cavity photon are usually a small perturbation.
Nevertheless, the mechanical effects due to the cavity field can become significant close to a structural instability. 

\begin{figure}[hbt]
\begin{center}
\includegraphics[width=0.36\textwidth]{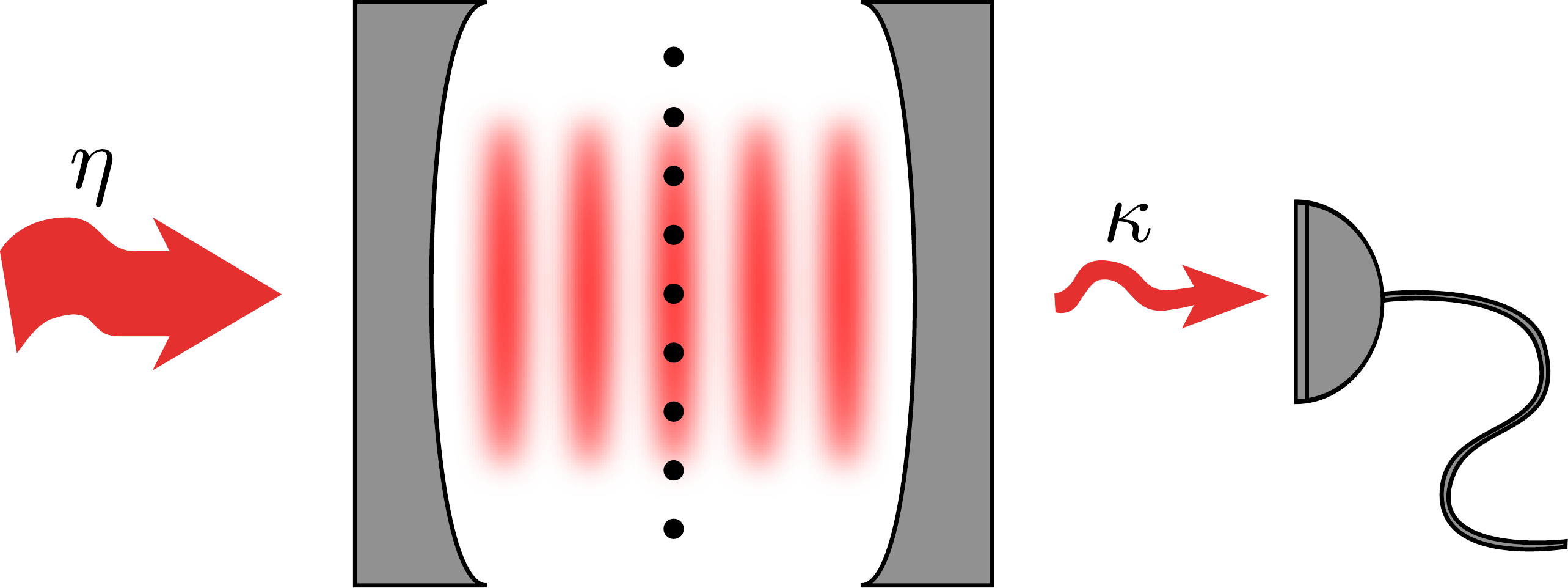}
\caption{\label{fig:system} (color online) The dipolar transitions of ions forming a chain strongly couple with one mode
of a high-finesse optical cavity. The photon-mediated interaction between the ions gives rise to multistable structural
configurations and to quantum correlations between photonic and mechanical fluctuations, which can be revealed by
measuring the intensity and spectrum of the field at the cavity output.}
\end{center}
\end{figure}

In Ref. \cite{Cormick_2012} we analyzed the equilibrium configurations and dynamics of an ion chain close to the
linear-zigzag instability \cite{Fishman_PRB_2008} and confined inside a high-finesse optical resonator. We showed that
linear and zigzag arrays are bistable for certain strengths of the laser pumping the cavity. For these regimes we argued
that the chain is cooled into one of the configurations by cavity-enhanced photon scattering, exhibiting steady-state
entanglement between photonic and vibrational fluctuations. In the present work we characterize the stationary state of
the ion chain in detail, focusing on the configuration in Fig.~\ref{fig:system}. We derive a comprehensive theoretical
model which allows us to find the regime in which the chain exhibits bistable structures. The steady state of cavity and
vibrational excitations is determined within a semiclassical approximation and the spectrum of light at the cavity
output is evaluated.  It is shown that motional and field fluctuations are 
entangled in the parameter regime where bistability is found. The entanglement is studied as a function of the pump
strength by means of the logarithmic negativity \cite{Vidal_Werner_PRA_2002}, and its signatures are  identified in the
spectrum of the light at the cavity output. 

This work is organized as follows. In Sec.~\ref{sec:system} the theoretical model of the system is introduced. In
Sec.~\ref{subsec:semiclassical} we describe the semiclassical treatment of the problem, and in Sec.~\ref{sec:mean
values} the equilibrium configurations of the crystal and cavity mode are analyzed as a function of the system
parameters. Section~\ref{sec:bistable} is devoted to the effect of the cavity field on the structural stability. The
equations governing the dynamics of the fluctuations of cavity field and motion about their equilibrium values are
derived in Sec.~\ref{sec:fluctuations}, and their solution is discussed in Sec.~\ref{sec:fluct evolution}. In
Sec.~\ref{sec:cooling} the stationary state of the fluctuations is characterized. The mean excitation number of the
different vibrational modes due to cavity cooling is reported in Sec.~\ref{sec:covariance}, the spectrum of the light at
the cavity output is given in Sec.~\ref{sec:spectrum}, and the entanglement between cavity and motional 
fluctuations is determined in Sec.~\ref{sec:entanglement}. The conclusions are drawn in Sec.~\ref{sec:conclusions},
while the appendices provide details of the calculations relevant to Sec.~\ref{sec:fluct evolution} and
Sec.~\ref{sec:spectrum}.

\section{Ion chains in a quantum potential} \label{sec:system}

\subsection{The model}

We consider $N$ ions of mass $m$ and charge $q$ confined by a harmonic potential. We assume for simplicity that the
motion of the ions is restricted to the $x-y$ plane, and denote the frequencies of the harmonic trap by $\omega_x$ and
$\omega_y$. In absence of coupling to other physical objects, their dynamics is determined by Hamiltonian 
\beq
\label{eq:Hmotion}
H_{\rm ions} = \!\sum_{j=1}^N \Bigg[\frac{\mathbf{p}_j^2}{2m} + V_{\rm trap} (\mathbf{r}_j) + \!\!\! \sum_{k=j+1}^N
\!\!\! V_{\rm Coul} (|\mathbf{r}_j-\mathbf{r}_k|)\Bigg] \,,
\eeq
where the operators $\mathbf{r}_j$ and $\mathbf{p}_j$ denote the position and momentum of the center of mass of the
$j$-th ion in the $x-y$ plane. $H_{\rm ions}$ includes the kinetic energy, the trap potential 
\beq
V_{\rm trap} (\mathbf{r}_j) = \frac{m}{2} (\omega_x^2 x_j^2 + \omega_y^2 y_j^2) \,,
\eeq
and the Coloumb repulsion
\beq
V_{\rm Coul}(|\mathbf{r}_j-\mathbf{r}_k|) = \frac{q^2}{4 \pi \epsilon_0} \frac{1}{\left| \mathbf{r}_j -
\mathbf{r}_k\right|} \,.
\eeq
At sufficiently low temperatures, which can be achieved by means of Doppler cooling, the ions crystallize at the
equilibrium positions determined by the trap potential and by the Coulomb repulsion \cite{Dubin_RMP99}. In this regime,
if $\omega_x$ exceeds a critical value $\omega_{crit}$ (which depends on $\omega_y$ and $N$), the ions form a linear
array along the $y$ direction \cite{Birkl_Kassner_Walther_Nat_1992,Birkl_PRL92,Dubin_PRL_1993,Morigi_2004}. For
$\omega_x$ below $\omega_{crit}$ but sufficiently close to it, the Coulomb repulsion pushes the ions away from the $y$
axis and the new equilibrium configuration has the geometry of a zigzag
\cite{Birkl_Kassner_Walther_Nat_1992,Birkl_PRL92,Dubin_PRL_1993,James_PRL_2000,Morigi_2004,Piacente_PRB_2004,
Fishman_PRB_2008}. 

In this article we assume that the ion chain is placed inside an optical resonator as shown in Fig. \ref{fig:system},
with $\omega_x>\omega_{crit}$.  A dipolar transition of the ions, formed by the ground state $\ket{g}$ and the excited
state $\ket{e}$ at transition frequency $\omega_0$, strongly couples with a cavity mode at frequency $\omega_c$, which
in turn is pumped by a laser at frequency $\omega_p$. The Hamiltonian describing the system dynamics, composed by the
cavity mode and the ions' internal and external degrees of freedom, reads
\beq
\label{eq:first Hamiltonian}
H = H_{\rm cav} + H_{\rm el}+ H_{\rm ions}  + H_{\rm JC}\,,
\eeq
where 
\beq
H_{\rm cav} = -\hbar \Delta_c a^\dagger a  - {\rm i} \hbar (\eta^* a - \eta a^\dagger)
\eeq
is the Hamiltonian for the cavity mode in the reference frame rotating with the laser frequency, with $a$ and
$a^{\dagger}$ the annihilation and creation operators of a cavity photon, $\Delta_c=\omega_p-\omega_c$ the detuning
between laser and cavity-mode frequency, while the frequency $\eta$ gives the strength of the pumping by the external
laser. The term
\beq
H_{\rm el} = -\hbar \Delta_0 \sum_{j=1}^N |e\rangle_j\langle e|
\eeq
accounts for the internal atomic dynamics, with $\Delta_0 = \omega_p -\omega_0$ the detuning of the pump from the
dipolar transition. Finally, the dipolar coupling between the cavity and the particles reads
\beq
H_{\rm JC} = \hbar \sum_{j=1}^N g(\mathbf{r}_j) (\sigma_j a^\dagger + \sigma^\dagger_j a) 
\eeq
and describes the absorption of a photon accompanied by the excitation of the atom $j$, given by the raising operator
$\sigma^\dagger_j=|e\rangle_j\langle g|$, and the emission with corresponding atomic de-excitation by the lowering
operator $\sigma_j$. Frequency $g(\mathbf{r}_j)$ is the coupling strength between the cavity mode and the ion at
position ${\bf r}_j$. It is modulated by the amplitude of the cavity field at the position of the atom and is here
assumed to take the form 
\beq
\label{eq:g}
g(\mathbf{r})= g_0 \cos(kx) e^{-y^2/(2w^2)}\,,
\eeq
where $g_0$ is the maximum value that the function $|g(\mathbf{r})|$ can take, $k$ is the modulus of the cavity wave
vector
$\mathbf{k}$ pointing along the $x$ axis, and $w$ is the width of the Gaussian transverse profile of the mode, with
$kw\gg1$. 

In addition, the coupling to the modes of the electromagnetic field external to the resonator gives rise to cavity
losses at rate 2$\kappa$ (we neglect here photon absorption at the cavity mirrors) and to spontaneous decay of the
atomic transition at rate $\gamma$. The incoherent effects are included in the dynamics using Heisenberg-Langevin
equations for the operators \cite{Gardiner_Collett_1985, Szirmai_etal_PRA_2010}. For instance, the equations of motion
for the operators $a$ and $\sigma_j$ read:
\begin{eqnarray}
\label{eq:a:0}
&&\dot{a}=\frac{1}{{\rm i}\hbar}[a,H]-\kappa a+\sqrt{2\kappa} \, a_{\rm in}(t)\\
&&\dot{\sigma_j}=\frac{1}{{\rm i}\hbar}[\sigma_j,H]-\frac{\gamma}{2}\sigma_j+\sqrt{\gamma} \, \sigma_{j,\rm
in}(t)\,,\label{eq:sigma}
\end{eqnarray}
with input noise $a_{\rm in}$ and $\sigma_{j,\rm in}(t)$ for field and atom, respectively. Denoting by $\zeta_{\rm
in}(t)=a_{\rm in}(t),\,\sigma_{j, \rm in}(t)$, the input noise operators have vanishing expectation value, $\langle
\zeta_{\rm in}(t)\rangle=0$, and correlations satisfying the relation
\beq \label{eq:a_input_comm}
\langle [\zeta_{\rm in}(t'), \zeta_{\rm in}^\dagger (t'')] \rangle = \delta (t'-t'') \,.
\eeq 
The state of the electromagnetic field outside the cavity is assumed to be thermal with mean number of photons $\bar
n(\omega_c)\approx 0$ at frequency $\omega_c$, such that 
\beq \label{eq:a-input}
\langle \zeta_{\rm in}^\dagger(t') \zeta_{\rm in} (t'') \rangle = 0\,.
\eeq
The equations of motion for $a^{\dagger}$ and  $\sigma_j^{\dagger}$ are the Hermitian conjugate of Eq. \eqref{eq:a:0}
and Eq. \eqref{eq:sigma}, respectively. The ions' motion is affected by other sources of noise such as patch potentials
in the electrodes. A specific model for the noise on the motion will be introduced in Section \ref{sec:fluctuations}.

\subsection{The quantum potential}

In this paper we characterize structural transitions induced by the coupling of the ion chain with the cavity field. We
consider the situation in which the linear chain in free space is stable, namely, $\omega_x>\omega_{crit}$, but close to
the mechanical instability, and we study the effect of the strong coupling with the cavity field on the structural
stability. We focus on the regime in which the detuning of the laser pump with respect to the atom, $\Delta_0$,
determines the fastest time scale, such that $|\Delta_0|\gg\gamma,\kappa,|\Delta_c|,g_0\sqrt{\bar n}$, with $\bar n$ the
mean intracavity photon number. In this case, the excited state of the atoms can be adiabatically eliminated from the
equations of motion for cavity and ions' external motion, leading to the effective interaction
\beq
H_{\rm int} = \hbar \, a^\dagger a \, U_0 (\mathbf{r}_1,\ldots ,\mathbf{r}_N) \,, \label{eq:Hint} 
\eeq
where $U_0$ reads
\beq
U_0(\mathbf{r}_1,\ldots ,\mathbf{r}_N)= \sum_{j=1}^N \frac{g^2(\mathbf{r}_j)}{\Delta_0} \,,
\label{eq:U:0}
\eeq
and weighs the nonlinear coupling between motion and cavity mode. 
Under this condition, the Hamiltonian governing the dynamics of cavity field and external degrees of freedom of
the ions now reads $H=H_{\rm ions}+H_{\rm cav}+H_{\rm int}$. Frequency $U_0$ is the shift of the cavity frequency due to
the ions inside the resonator, and conversely it is the mechanical potential exerted on these ions by a single cavity
photon \cite{Domokos_Ritsch_2002,Maschler_Ritsch_PRL_2005, Larson_et_al_PRL_2008}. Hamiltonian $H_{\rm int}$ thus can be
interpreted as a quantum potential for the atoms, being its amplitude dependent on the number of intracavity photons.
This term gives rise to mechanical effects that, for strong coupling, can be significant even at the single-photon
level. 
Parameter $U_0$ also scales photon losses due to spontaneous emission. In fact, for a given set of ions' positions the
cavity decay rate reads $\kappa_{\rm eff}=\kappa + U_0 \gamma /(2\Delta_0)$. In the following we shall restrict to the
regime in which the cavity-ion interaction is mainly dispersive and spontaneous emission can be thus neglected.

\section{Stationary state} \label{sec:stationary}

In this section we determine the stationary state of the coupled system, in the regime in which the relevant degrees of
freedom are the external motion of the ions and the cavity field, which are coupled via the effective potential in Eq.
\eqref{eq:Hint}.

\subsection{The semiclassical limit} \label{subsec:semiclassical}

We perform the study in the semiclassical limit, assuming that the fluctuations about the mean values of field and
atomic variables are sufficiently small to justify the treatment.  To this aim, we decompose the operators as a sum of
mean values and fluctuations according to the prescription 
\beq
\label{eq:mean:delta}
\begin{array}{l}
 a = \bar a + \delta a \,,\\
 \mathbf{r}_j = \bar{\mathbf{r}}_j + \delta \mathbf{r}_j \,,\\ 
 \mathbf{p}_j = \bar{\mathbf{p}}_j + \delta \mathbf{p}_j \,,
\end{array}
\eeq
where $\langle a \rangle= \bar a$,   $\langle \mathbf{r}_j \rangle=  \bar{\mathbf{r}}_j $, and $\langle \mathbf{p}_j
\rangle=  \bar{\mathbf{p}}_j$, while the expectation value of the fluctuations $\delta a, \delta \mathbf{r}_j, \delta
\mathbf{p}_j$ vanishes. The mean values satisfy the equations of motion
\begin{align}
 & \frac{\partial~}{\partial t} \bar a = ( {\rm i} \Delta_{\rm eff} - \kappa) \bar a + \eta \,, \label{eq:a eq}\\
 & \frac{\partial~}{\partial t} \bar{\mathbf{r}}_j = \frac{\bar{\mathbf{p}}_j}{m} \,, \label{eq:r eq} \\
 & \frac{\partial~}{\partial t} \bar{\mathbf{p}}_j = - \grad_j H_{\rm ions} - \hbar |\bar a|^2 \grad_j U_0 \,,
\label{eq:p eq}
\end{align}
with $\grad_j$ the gradient with respect to the spatial coordinates of the $j$-th particle (evaluated at the equilibrium
positions $\bar{\mathbf{r}}_1,\ldots ,\bar{\mathbf{r}}_N$), while
\beq
\label{Delta:eff}
\Delta_{\rm eff}=\Delta_c-U_0(\bar{\mathbf{r}}_1,\ldots ,\bar{\mathbf{r}}_N)\,.
\eeq
In order to determine the classical equilibrium values we require that the quantities $\bar a$, $\bar{\mathbf{r}}_j$ and
$\bar{\mathbf{p}}_j$ correspond to stationary solutions of the dynamical equations, namely $\partial_t \bar a=0$, $
\partial_t\bar{\mathbf{r}}_j =0$, and $\partial_t\bar{\mathbf{p}}_j =0$. 

The coupled dynamics of the quantum fluctuations of field and motion are governed by the Heisenberg-Langevin equations
\cite{Gardiner_Collett_1985, Szirmai_etal_PRA_2010}, which are found substituting the decomposition
\eqref{eq:mean:delta} into Eq. \eqref{eq:a:0} and into the Heisenberg equations of motion for the center-of-mass
variables, and using that the mean values are the stationary solutions. The equations read 
\begin{align}
\label{eq:a fluct}
\delta\dot{a} &= ({\rm i} \Delta_{\rm eff} - \kappa) \delta a - {\rm i} \bar a \sum_k (\delta \mathbf{r}_k \grad_k) U_0
+ \sqrt{2\kappa}\, a_{\rm in}\,, \\
\label{eq:r fluct}
\dot {\delta \mathbf{r}_j} &= \frac{\delta \mathbf{p}_j}{m}\,, \phantom{\sum_k}\\
\label{eq:p fluct}
\dot {\delta \mathbf{p}_j} &= - \sum_k (\delta \mathbf{r}_k \grad_k) \left( \grad_j H_{\rm ions} + \hbar |\bar a|^2
\grad_j U_0 \right) \nonumber \\
& \quad \, - \hbar \left({\bar a}^* \delta a + \bar a \delta a^\dagger\right) \grad_j U_0 \,,
\end{align}
where the derivatives in the expressions above are evaluated at the equilibrium positions. With no loss of generality,
we choose the phase of $\eta$ such that $\bar a$ is real.

\subsection{Equilibrium configurations}  \label{sec:mean values}

We first focus on the equilibrium configuration found by setting Eqs. \eqref{eq:a eq}-\eqref{eq:p eq} equal to zero.
From Eq. (\ref{eq:a eq}) the equilibrium value of the cavity field amplitude reads
\beq
\label{bar:a}
\bar a=\frac{\eta}{\kappa-{\rm i}\Delta_{\rm eff}} \,,
\eeq
and depends on the positions of the ions. Setting Eq.~(\ref{eq:r eq}) to zero gives $\bar{\mathbf{p}}_j=0$, and finally
from Eq. (\ref{eq:p eq}) one finds that the equilibrium positions of the ions must be minima of an effective potential
of the form
\beq \label{eq:Vtot}
V_{\rm tot} = V_{\rm trap} + V_{\rm Coul} + V_{\rm eff} \,,
\eeq 
where the term
\beq \label{eq:eff opt potential}
V_{\rm eff} = \frac{\hbar |\eta|^2}{\kappa} \arctan \left(-\frac{\Delta_{\rm eff}}{\kappa} \right)
\eeq
is due to the coupling with the cavity field. The back-action of the cavity field on the ions, in particular, enters in
Eq. \eqref{eq:eff opt potential} via the parameter $\Delta_{\rm eff}$, Eq. \eqref{Delta:eff}, and scales with $U_0$.
Whether cavity back-action is relevant to the dynamics can be determined by means of the dimensionless parameter 
\beq
\label{Coop}
\mathcal C = \frac{g_0^2N_{\rm eff}}{\kappa |\Delta_0|} \,,
\eeq
where $N_{\rm eff}$ is the effective number of ions that couple to the field mode, defined as:
\beq
N_{\rm eff} = \sum_{j=1}^N e^{-\bar y_j^2/w^2}.
\eeq
Parameter \eqref{Coop} can be identified with the cooperativity  \cite{Kimble}.  For $\mathcal{C}\gtrsim1$ the potential
in Eq. \eqref{eq:eff opt potential} gives rise to an effective long-range force between the ions. On the other hand, in
the limit $\mathcal C \ll 1$ the effect of the ions on the field is negligible. Then, the mean value of the cavity field
is determined by external parameters, so that $\bar a = \eta/(\kappa-i\Delta_c)$ and Eq. (\ref{eq:Hint}) is well
approximated by a classical potential of the form:
\beq
\label{eq:Vopt}
V_{\rm opt} = \frac{\hbar \, |\eta|^2}{\kappa^2+\Delta_c^2} \, \sum_{j=1}^N \frac{g^2(\mathbf{r}_j)}{\Delta_0}\,,
\eeq
which overlaps to the trap and Coulomb potential. 

In this paper we analyze the case when the string is orthogonal to the cavity-mode wave vector, as in Fig.
\ref{fig:system}, corresponding to the limit $\omega_x\gg\omega_y$. The formalism developed so far, however, is
valid for any structure that the ions can take. It can be applied, hence, also for the
specific case in which the ions form a string which is parallel to the cavity axis. In our model, this case corresponds
to 
setting $\omega_y\gg\omega_x$ so that the equilibrium configuration has values $y_j=0$ for $j=1,,\ldots, N$. For small
cooperativity, $\mathcal C\ll 1$, the equilibrium positions along $x$ correspond to minima of the total potential:
\begin{multline}
V_{\rm FKM} = \sum_{j=1}^N \Bigg[ \frac{m}{2} \omega_x^2 x_j^2 + \sum_{k=j+1}^N \frac{q^2}{4 \pi \epsilon_0}
\frac{1}{\left|x_j - x_k\right|}\\
+ \frac{\hbar \, |\eta|^2}{\kappa^2+\Delta_c^2} \frac{g_0^2}{\Delta_0}\, \cos^2(kx_j) \Bigg] \,,
\end{multline}
and can be mapped to a Frenkel-Kontorova model \cite{Garcia-Mata_Zhirov_Shepelyansky_2007, Pruttivarasin_NJP_2011}. Upon
varying the depth of the optical potential, this model exhibits a classical transition between sliding and pinned
phases, which has been proposed to study friction \cite{Benassi_Vanossi_Tosatti_2011}, and a quantum transition between
a pinned instanton glass and a sliding phonon gas \cite{Garcia-Mata_Zhirov_Shepelyansky_2007}. If, instead, the
effective cooperativity $\mathcal C$ is larger than unity, the term associated to the optical potential introduces also
an additional, infinitely-ranged interaction between the particles, giving rise to dynamics which are largely unexplored
to date.

\subsection{Multistability}
\label{sec:bistable}

We shall assume that the string is orthogonal to the cavity-mode wave vector, as in Fig. \ref{fig:system}. Such a
scenario can be realized, for instance, with
the setups of Refs. \cite{Keller_etal_Nat_2004, Stute_etal_Nat_2012}. 
In this case
the optical potential generates a transverse force, which is symmetric about the chain axis when the chain is at a node
or antinode of the cavity standing wave. 

We first consider small cooperativities, $\mathcal C\ll 1$. If all ions are illuminated by the cavity mode, close to the
linear-zigzag mechanical instability the optical potential shifts the critical value of the transverse trap frequency
$\omega_x$ with respect to the free-space value $\omega_{crit}$. If only a part of the chain is illuminated, at a value
$\omega_x>\omega_{crit}$ a local structural distortion will appear with the form of a zigzag chain. Let us assume that
the equilibrium positions of the ions in the linear array are located at an antinode of the cavity mode. For
blue-detuned pumps, with $\Delta_0>0$, the light field pushes the particles away from the antinode and a mechanical
instability thus appears at frequencies $\omega_x$ larger than $\omega_{crit}$, while a red-detuned pump field has the
opposite effect. We note that these generic considerations, which are valid for the secular potential of a Paul trap,
can also be applied when the micromotion is taken into account. In fact, the plane 
containing the zigzag array can be chosen to be such that micromotion is in the direction perpendicular to the plane and
therefore has no effect on the problem under study.

\subsubsection{Multistable structures}

\begin{figure}[hbt]
\begin{center}
\includegraphics[width=0.42\textwidth]{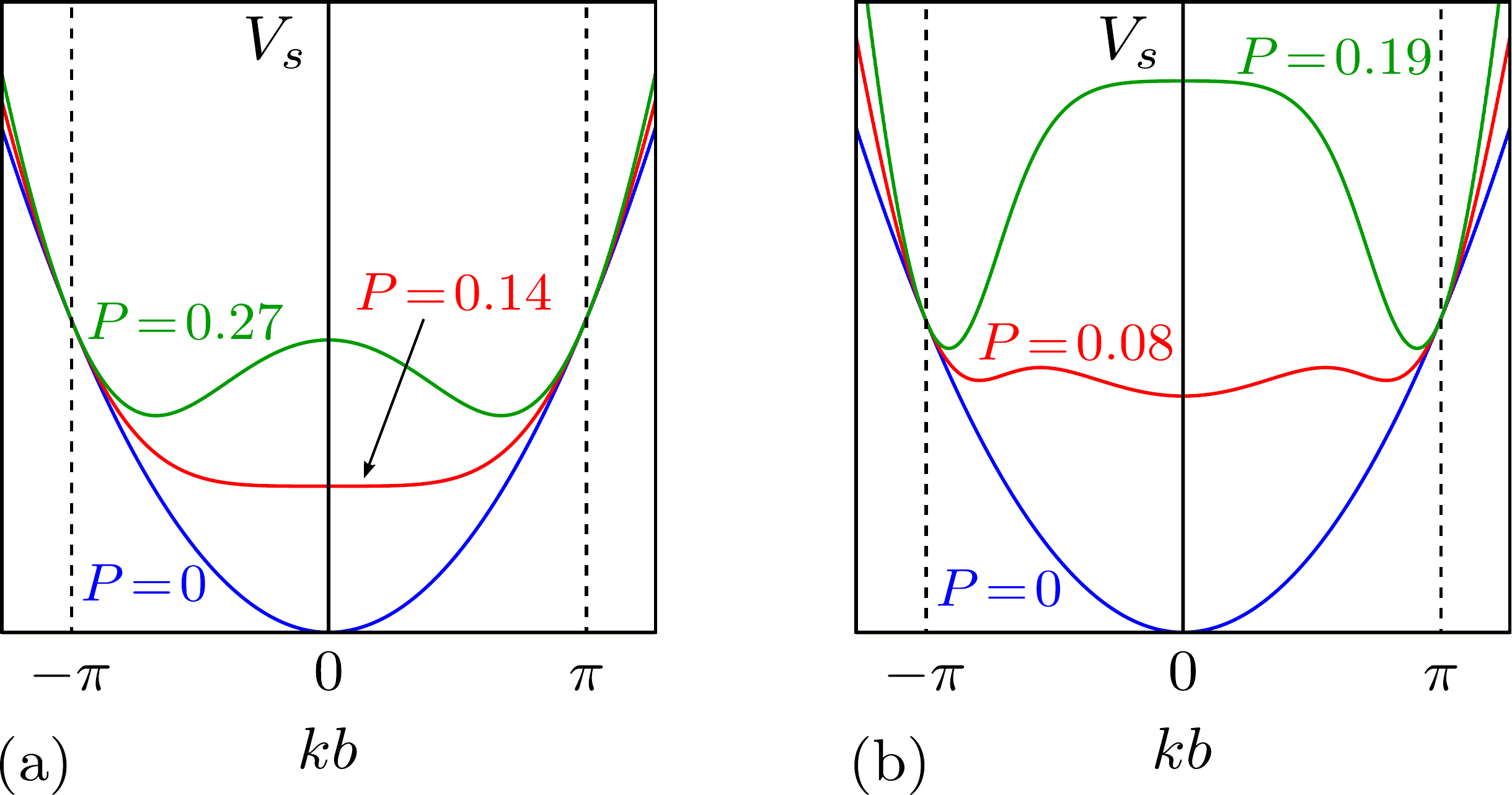}
\end{center}
\caption{\label{fig:infinite chain} (color online) Potential $V_s$ of the zigzag mode, Eq.~\eqref{eq:Vs}, as a function
of the zigzag amplitude $b$ (in units of $1/k$) for different pumping powers $P$ (in units of
$P_0=\kappa\omega_x^2/\omega_R$). The dashed lines indicate the nodes of the cavity mode at $kb=\pm\pi$. The
cooperativity is (a) $\mathcal C=0.5$ and (b) \mbox{$\mathcal C=3$}. The coupling of the ions with the field is assumed
to be
homogeneous and $\vartheta = N \omega_s^2/\omega_x^2 = 0.22$. 
}
\end{figure}

The dynamical behaviour of the system is significantly modified at large cooperativities, $\mathcal C\gtrsim1$, where
the cavity-mediated interaction between the ions becomes relevant. This property introduces an additional nonlinearity.
For definiteness, from now on we take $\Delta_c=0$, $\Delta_0>0$, and we restrict to the case when $\omega_x >
\omega_{crit}$ so that in the absence of pumping the particles form a linear array along $x=0$, coinciding with an
antinode of the cavity field. Therefore, the effective detuning $|\Delta_{\rm eff}|$ is maximum when the chain is in the
linear configuration. For a fixed pumping strength $\eta\neq0$, the intracavity field intensity increases if the
equilibrium positions are shifted to a zigzag. This optical nonlinearity can lead to bistable linear and zigzag
configurations, in contrast to the continuous linear-zigzag transition in free space. 
This can be better understood by analyzing the effective potential $V_s$ of the zigzag mode. We start from the knowledge
that the zigzag mode is the soft mode of the linear-zigzag transition in free space and that an effective, Landau
potential for the soft mode can be derived, whose equilibrium solutions are either the linear or the zigzag
configurations  \cite{Fishman_PRB_2008}. We then study how the effective potential is modified by the coupling with the
cavity, thereby discarding the coupling with the other vibrational modes that may arise due to the cavity field. We
first consider the simplest limit assuming that the ions of the linear array are equidistant. In this case the zigzag
mode amplitude reads $x_s = \sum_j (-1)^j x_j/\sqrt{N}$, while the zigzag amplitude, equal to twice the transverse
equilibrium
displacement of the ions from the $x=0$ axis, is $b = 2x_s/\sqrt{N}$. Denoting by
$\omega_s=\sqrt{\omega_x^2-\omega_{crit}^2}$ the frequency of the soft mode in free space and
above 
the critical 
point, the potential $V_s$ can be written as
\beq 
\label{eq:Vs}
V_s = \frac{m\omega_x^2}{k^2} \left\{ \frac{\vartheta}{2} \left(\frac{kb}{2}\right)^2 + 2 P \arctan \left[\mathcal C
\cos^2\left(\frac{kb}{2}\right) \right] \right\} \,,
\eeq 
where we assumed that the ions are uniformly illuminated by the cavity field. 
The first term on the right-hand side of Eq. (\ref{eq:Vs}) describes the harmonic potential for the soft mode in free
space, with $\vartheta = N \omega_s^2/\omega_x^2$. The second term corresponds to the optical potential, proportional to
the dimensionless power $P = |\eta|^2\omega_R/(\kappa\omega_x^2)$, with $\omega_R=\hbar k^2/(2m)$ the recoil frequency.
The form of the potential $V_s$ is shown in Fig. \ref{fig:infinite chain} for different values of $\mathcal C$ and $P$.
The stability of the linear configuration is determined by $\mathcal C$ and by the ratio $P/\vartheta$ and corresponds
to the presence of a minimum at $b=0$. For $P=0$ the cavity mode is in the vacuum state and the linear array is stable.
The soft mode becomes unstable when the optical power is increased above the threshold value $P = \vartheta (1+\mathcal
C^2)/(4\mathcal C)$. For large cooperativities, there are parameter regimes for which both linear and zigzag
configurations are stable, as illustrated by the appearance of three minima 
in Fig.~\ref{fig:infinite chain}(b) for intermediate values of the pump power. 

These results are based on the assumption that the interparticle distance is uniform. Such model can be realized in a
ring trap  \cite{Birkl_Kassner_Walther_Nat_1992} or in a multipolar radiofrequency trap  \cite{Okada_etal_PRA_2007,
Champenois_etal_PRA_2010}, and it approximates a chain in an axial anharmonic trap \cite{Lin_EPL_2009}. For a chain in a
linear Paul trap, however, the interparticle distance varies \cite{Dubin:1997}, and neither the coupling of the ions to
the cavity nor the corresponding amplitude of the soft mode are uniform along the chain \cite{DeChiara_NJP:2010}. Figure
\ref{fig:hysteresis}(a) displays $V_s$ for a chain of 60 ions in a harmonic trap, where the central region of the chain
couples to the cavity mode with $N_{\rm eff}\sim 5.7$. Here, for certain values of the pumping strength the potential
$V_s$ also exhibits three minima, corresponding to stable linear and zigzag arrays. We note that close to the
instability the equilibrium positions in the zigzag configuration are very 
similar in geometry to the soft mode of the linear chain.

\subsubsection{Intensity of the light at the cavity output}

\begin{figure}[hbt]
\begin{center}
\includegraphics[width=0.45\textwidth]{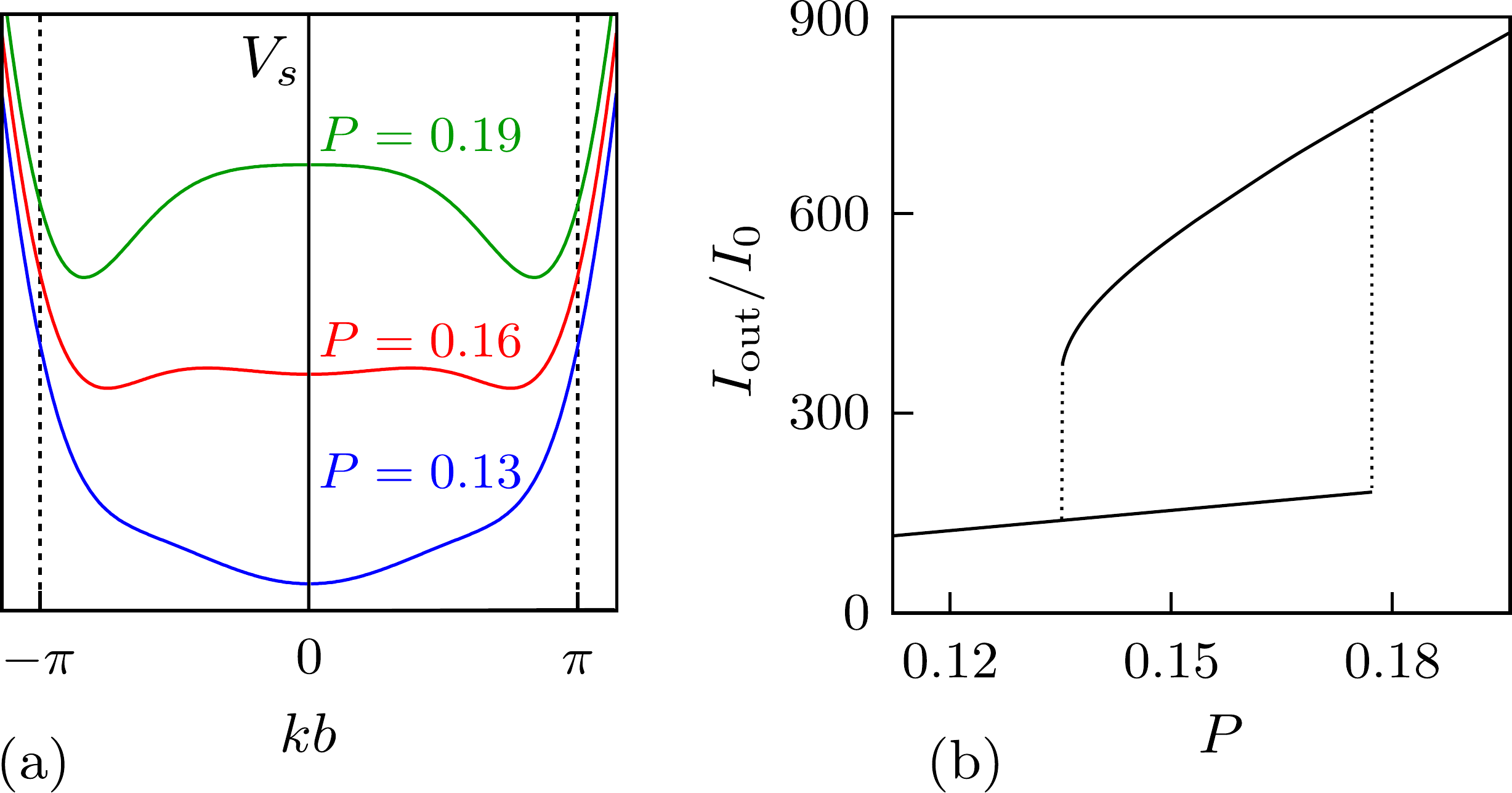}
\end{center}
\caption{\label{fig:hysteresis} (color online) (a) Same as Fig. \ref{fig:infinite chain} but for a chain in a linear
Paul trap. A chain of 60 $^{40}$Ca$^+$ ions is taken with interparticle distance 4.3 $\mu$m in the central region. The
ions are coupled to a cavity mode with wavelength 866 nm and transverse width $w=14\,\mu$m ($N_{\rm eff}\simeq 5.7$).
(b) Intensity $I_{\rm out}$ at the cavity output as a function of $P$. Here, the intensity is in units of $I_0 = I_{\rm
out}(P_0)$, and for a configuration minimizing $V_{\rm tot}$, numerically found using linear and zigzag chains as
initial guesses. The other parameters are $\omega_y =2\pi~\times$ 0.1~MHz, $\omega_x = 2\pi~\times$ 2.26~MHz (the
critical value is $\omega_{crit}\simeq 2\pi~\times$ 2.216~MHz), $\Delta_c=0$, $\Delta_0=2\pi~\times$ 500~MHz, 
$\kappa=2\pi~\times$ 0.5 MHz, $g_0=2\pi~\times$ 9.4 MHz, $\gamma=2\pi~\times$ 10~MHz, resulting in $\mathcal C = 2$.}
\end{figure}

The bistable behaviour can be detected by monitoring the mean value of the intensity $I_{\rm out}$ of the field at the
cavity output. This is found from the mean value of the field at the cavity output, $a_{\rm out}=\sqrt{2\kappa}a+a_{\rm
in}$ and reads
\beq
I_{\rm out}=2\kappa |\bar a|^2\,.
\eeq
Figure \ref{fig:hysteresis}(b) displays $I_{\rm out}$ as a function of the pump intensity. The lower branch corresponds
to the light at the cavity output when the ions form a linear array, and the upper branch to a zigzag array.
The presence of two different values of the output intensity for a given pump power indicates bistable equilibrium
configurations of the ions' structure. 

\subsubsection{Discussion}

The bistability we predict is a consequence of the dependence of the optical potential on the positions of the atoms
within the standing-wave field. In the thermodynamic limit, if the region of the chain interacting with the cavity mode
is finite, the effect of this coupling is a localized defect in the chain. For a finite system, nevertheless, forces
acting on few ions can generate arrays close to zigzag configurations due to the long-range Coulomb repulsion
\cite{Baltrusch_etal_PRA_2011, Li_Lesanovsky_PRL_2012}.

The observed bistable behaviour shares several analogous features with textbook optical bistability
\cite{Lugiato}. The nonlinear dynamics here studied, however, are solely due to the interplay between the dispersive
optomechanical coupling of the ions' external motion with the cavity field and the long-range Coulomb interaction, for
configurations in which the net force over the ions vanishes. The nonlinear dependence on the ion positions enters
via the Coulomb repulsion and the detuning $\Delta_{\rm eff}$, Eq. \eqref{Delta:eff}, in the cavity-induced term of the
total potential in Eq. \eqref{eq:Vtot}. Therefore, bistability in this case
is not necessarily found for the configurations where the mean amplitude of the field in Eq. \eqref{bar:a} exhibits
local maxima, which would correspond to configurations where $\Delta_{\rm eff}$ vanishes. This is a major difference
with bistability effects in the optomechanical coupling between neutral atoms and cavity 
fields \cite{Hemmerich,StamperKurn,Ritter,Larson:2009}, where the strong coupling with the cavity gives rise to the most
relevant nonlinearity. We note that optomechanical bistability, for media constituted by neutral quantum gases, can be
associated with different quantum phases of ultracold atomic gases of bosons \cite{Larson_et_al_PRL_2008,Meystre}.
Analogously, in our case bistability is associated with different classical structures. 

\subsection{Quantum fluctuations} \label{sec:fluctuations}

We now analyze the dynamics of the fluctuations of the system observables about the equilibrium values close to the
linear-zigzag instability, where multistability due to the strong coupling with the cavity mode is observed. We 
restrict to the case when these fluctuations are sufficiently small that the system admits a linearized description.
This requires the validity of the Lamb-Dicke approximation \cite{Stenholm_1986}, namely, the displacements from the
equilibrium positions have to be smaller than the light wavelength (therefore, this treatment is not valid in the very
close vicinity of the structural transition or at high temperatures).

For convenience we introduce the normal modes of the crystal, that characterize the dynamics of the ions when the
coupling with the quantum fluctuations of the cavity field can be neglected. We hence write the displacements of the
ions as: 
\beq
\delta x_j = \sum_n M_{jn}^{(x)} q_n \, , \quad \delta y_j = \sum_n M_{jn}^{(y)} q_n, \\
\eeq
with $M_{jn}^{(x,y)}$ the element of the orthogonal matrix relating the local coordinates $\delta x_j$, $\delta y_j$
with the normal-mode coordinates $q_n$. The normal-mode coordinates diagonalize Eqs. (\ref{eq:r fluct})-(\ref{eq:p
fluct}) when the cavity fluctuations $\delta a$ are set to zero. We note that this is not equivalent to setting the
cavity field to zero, since we define the normal modes taking into account the optical potential determined by the mean
value of the field. 

We denote by $b_n$ and $b_n^\dagger$ the bosonic operators annihilating and creating, respectively, an energy quantum
(phonon) of the normal mode at frequency $\omega_n$. They are defined through the equations
$q_n=\sqrt{\hbar/(m\omega_n)} Q_n$ and $p_n=\sqrt{\hbar m\omega_n}P_n$, with 
\beq
\label{eq:motion quad}
Q_n = \frac{b_n+b_n^\dagger}{\sqrt{2}}, \quad~ P_n = {\rm i}\frac{b_n^\dagger-b_n}{\sqrt{2}}\,.
\eeq
The dynamical equations for motional modes and photonic fluctuations take then the form:
\begin{align}
\label{eq:a}
\!\!\!& \delta \dot a = ({\rm i} \Delta_{\rm eff} - \kappa) \delta a - {\rm i} \bar a \sum_n c_n (b_n + b_n^\dagger) +
\sqrt{2\kappa}\, a_{\rm in}\,, \\
\label{eq:b}
\!\!\!& \dot b_n = - ({\rm i} \omega_n + \Gamma_n) b_n - {\rm i} \bar a c_n (\delta a + \delta a^\dagger) +
\sqrt{2\Gamma_n} \, b_{{\rm in}, n}\,,
\end{align}
which also includes the effect of a noise source heating the vibrational mode $n$ at rate $\Gamma_n$. The corresponding
Langevin force is described by the input noise operator $b_{{\rm in}, n}$, with $\langle b_{{\rm in},n} \rangle = 0$ and
\cite{Gardiner_Collett_1985}:
\beq \label{eq:b-input}
\langle b_{{\rm in}, n}^\dagger(t') \, b_{{\rm in}, n'} (t'') \rangle = \bar N_n \, \delta_{n n'} \, \delta(t'-t'')\,.
\eeq
Here, $\bar N_n = \bar N (\omega_n)$ is the mean excitation number of an oscillator of frequency $\omega_n$ at the 
temperature of the considered environment \footnote{The other non-vanishing correlations can be obtained from the
commutation relations:
\beq \nonumber 
\langle [b_{{\rm in}, n}(t'), b_{{\rm in}, n'}^\dagger (t'')] \rangle = \delta_{nn'} \delta (t'-t''). \quad
\eeq}.
These terms introduce a simple model simulating the heating observed in an ion trap \cite{Haeffner_Roos_Blatt_PRep_2008,
Wineland_etal_NIST_1998, James_PRL98, Henkel_1999}. 

The coefficients $c_n$ in Eq. \eqref{eq:a}-\eqref{eq:b} give the coupling strength between motional and cavity field
fluctuations. They read
\beq
c_n = \sqrt{\frac{\hbar}{2m\omega_n}} \sum_j \left[M_{jn}^{(x)} \partial_{x_j} + M_{jn}^{(y)} \partial_{y_j} \right]
U_0\,, \label{eq:cavity-motion-coupling}
\eeq
where the derivatives are evaluated at the equilibrium positions $\bar{\mathbf{r}}_j$. They vanish when the equilibrium
positions $\bar{\mathbf{r}}_j$ are located at field nodes. If the particles are instead at antinodes, the coupling is
determined by the derivatives in $y$ direction, which are assumed to be much smaller than those along $x$ (since
$kw\gg1$). As we assume that when the ions form a linear chain they lie at an antinode, the coupling between vibrations
and field fluctuations is much stronger in the zigzag configuration than for the linear array.

\subsection{Solution of the linearized coupled evolution of fluctuations} \label{sec:fluct evolution}

We now solve the linear inhomogeneous system of differential equations (\ref{eq:a})-(\ref{eq:b}). 
Following a standard procedure, we look for the solution of the homogeneous system and for a particular solution of the
inhomogeneous equations \cite{Vitali_2008,Tufarelli_etal_2012}. We first introduce dimensionless quadrature operators
for field and ions' motion, 
\beq
 X_a = \begin{pmatrix} Q_a\\P_a \end{pmatrix}\,,
 \quad 
 X_n = \begin{pmatrix} Q_n\\P_n \end{pmatrix}\,,
\eeq
with the field quadratures
\beq
\label{eq:field quad}
Q_a = \frac{\delta a + \delta a^\dagger}{\sqrt{2}}, \quad P_a = - {\rm i} \: \frac{\delta a - \delta
a^\dagger}{\sqrt{2}}\ ,
\eeq
while $Q_n$ and $P_n$ are defined in Eq. \eqref{eq:motion quad}. We then arrange them together in a column vector,
\beq 
\label{eq:Xdef}
\vec X = \begin{pmatrix} X_a \\ X_1 \\ X_2 \\ \vdots \\ X_N \end{pmatrix} \, ,
\eeq
and rewrite Eqs. (\ref{eq:a})-(\ref{eq:b}) in the compact form
\beq
\label{eq:linear compact}
\frac{d\vec X}{dt} = M \vec X + \vec X_{\rm in}(t) \,.
\eeq
Here, $\vec X_{\rm in}$ contains the input noise operators, while
\beq 
\label{eq:M}
M = 
\begin{pmatrix} 
M_a & A_1 & A_2 & \ldots & A_N \\
A_1 & M_1 & 0 & \ldots & 0 \\
A_2 & 0 & M_2 & \dots & 0 \\
\vdots & \vdots & \vdots & \ddots & \vdots \\
A_N & 0 & 0 & \ldots & M_N
\end{pmatrix} \,,                                                 
\eeq
where
\begin{align}
 & M_a = -\kappa \, \mathbb{I} - {\rm i} \, \Delta_{\rm eff} \, \sigma_y\,,\\
 & M_n = -\Gamma_n \, \mathbb{I} + {\rm i} \, \omega_n \, \sigma_y\,,\\
 & A_n = -\bar a \, c_n (\sigma_x - {\rm i} \, \sigma_y) \,,
\end{align}
and $\sigma_{x,y,z}$ the Pauli matrices.

For each specific set of coefficients, the solution can be found by diagonalizing $M$. The time evolution of the
eigenvector of $M$ with eigenvalue $\lambda$ will follow an exponential function, which is decreasing (increasing) if
the real part $\Re(\lambda)$ is negative (positive), and oscillating if $\lambda$ is purely imaginary. Therefore, the
stability of the system requires that there are no eigenvalues with positive real parts \footnote{When the system is not
diagonalizable, the equations can be solved by means of the Jordan form of $M$, and the general solution is a sum of
exponentials multiplied by polynomials in time. The stability is still determined by the real parts of the eigenvalues
(except in the pathological case of a non-diagonalizable subspace with purely imaginary $\lambda$: this corresponds to
an unstable solution that grows polynomially in time). Criteria for the stability of similar systems of equations can be
found for instance in \cite{Vitali_etal_PRL_2007, Paternostro_etal_NJP_2006}.}. For a stable system, the values of
$-\Re(\lambda)$ determine the rates with which excitations decay. In particular, if
the cavity is the only environment the trace of $M$ is $-2\kappa$, which equals the sum of all the eigenvalues of the
system. This shows that the total cooling rate due to the coupling with the cavity cannot exceed $2\kappa$, so that the
cooling of a higher number of modes can only come at the price of lower cooling rates for each mode.

The eigenvalues $\lambda$ of $M$ satisfy the equation
\beq
\label{eq:eigenvalues}
\Delta_{\rm eff}^2 + (\kappa+\lambda)^2 = -4 \Delta \bar a^2 \sum_n \frac{c_n^2 \omega_n}{\omega_n^2 +
(\lambda+\Gamma_n)^2} \,,
\eeq
whose derivation is given in Appendix \ref{ap:stability}. The solutions are found numerically, but some of their
properties can be inferred from the form of the equations like, for instance, that complex eigenvalues come always in
complex-conjugated pairs. In Appendix \ref{ap:stability} it is also shown that, if $\Delta_{\rm eff}<0$ (the case where
multistable configurations are found in Sec. \ref{sec:bistable}), the condition of stability corresponds to the
inequality 
\beq \label{eq:stability}
\Delta_{\rm eff}^2+ \kappa^2 \geq -4\Delta_{\rm eff} \, \bar a^2 \sum_n \frac{c_n^2 \omega_n}{\omega_n^2 +
\Gamma_n^2}\,,
\eeq
which has been found assuming that the equilibrium positions $\bar{\mathbf{r}}_j$ correspond to minima of the effective
potential as determined by Eqs. (\ref{eq:Vtot})-(\ref{eq:eff opt potential}), so $\omega_n>0$ for all normal modes of
the crystal. 

For the cases we analyze in this paper the system of equations is diagonalizable, so that $M = T D T^{-1}$ with $T$ a
non-singular matrix and $D$ the diagonal matrix containing the eigenvalues. The solution of Eq. (\ref{eq:linear
compact}) takes the form
\beq \label{eq:general solution}
\vec X (t) = T e^{D(t-t_0)} T^{-1} \vec X (t_0) + \int_{t_0}^t dt' \, T e^{D(t-t')} T^{-1} \vec X_{\rm in}(t')\,,
\eeq
and will be used in order to evaluate the stationary state of the system.

\section{Cooling and stationary entanglement} \label{sec:cooling}

\begin{figure}[htb]
 \includegraphics[width=\columnwidth]{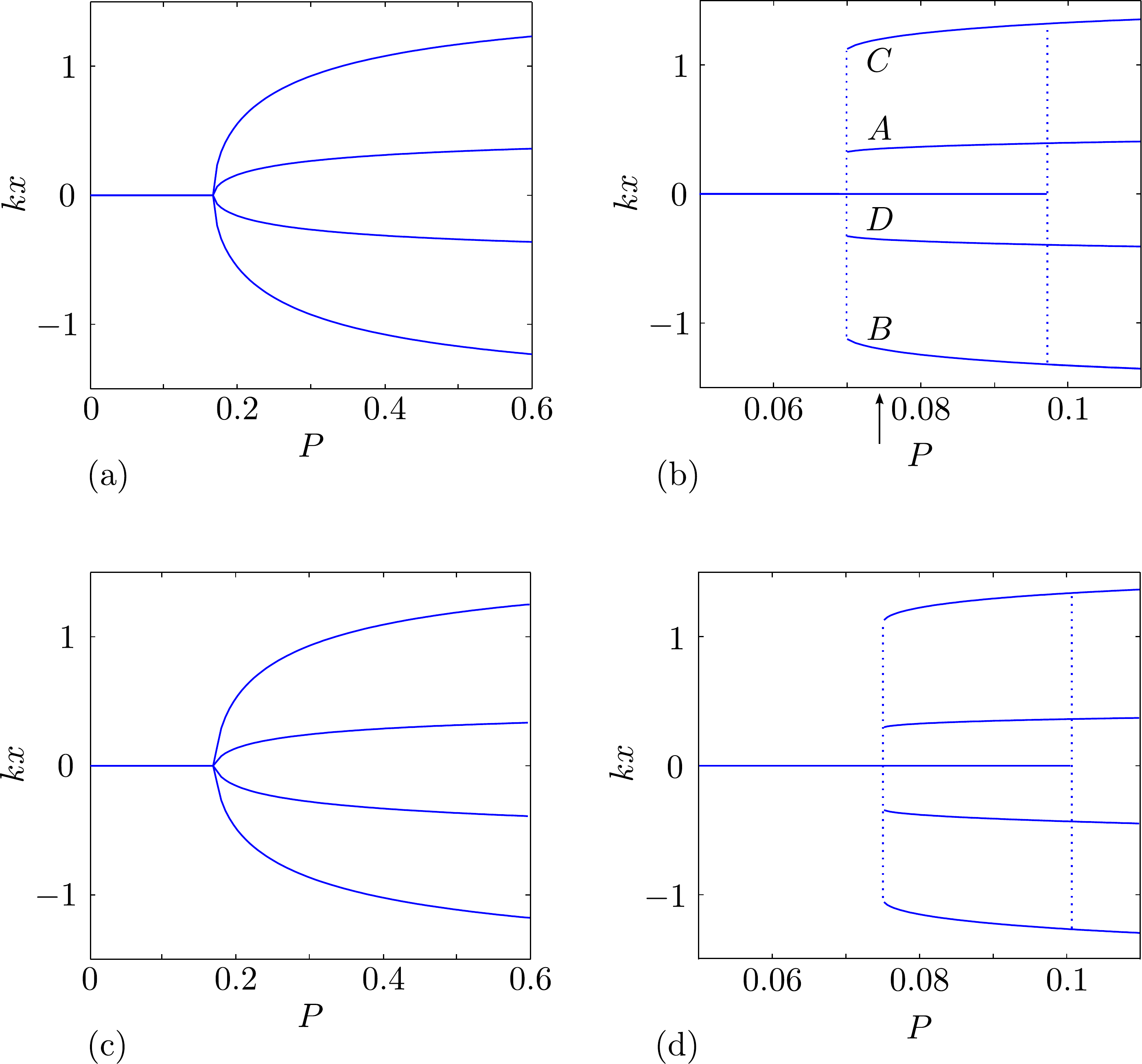}
 \caption{\label{fig:4ions} (color online) Equilibrium configurations in a chain of 4 ions as a function of the pump
power $P$. The plots display the value of the $x$ coordinate (in units of $1/k$) of each of the ions as a function of
$P$ (in units of $P_0=\kappa\omega_x^2/\omega_R$). The cooperativity is $\mathcal C = 0.3$ (left panels) and $\mathcal C
= 3$ (right panels).  The vertical dotted lines in the right panels indicate the limits of the bistability region. In
(a) and (b) the chain is centered with respect to the cavity axis, while in (c) and (d) it is displaced by 1.1~$\mu$m in
$y$ direction with respect to the cavity center. The arrow in (b) indicates the value of $P$ for the eigenmodes shown in
Fig. \ref{fig:modes}. The ions are labelled from $A$ to $D$, with $A$ and $D$ corresponding to the chain edges. The
other parameters are $\Delta_c=0$, $\Delta_0=2\pi~\times$ 500 MHz,  $\kappa=2\pi~\times$ 0.5 MHz, $\gamma=2\pi~\times$
10 MHz, $\omega_y = \kappa = 2\pi~\times$ 1 MHz, $\omega_x = 2\pi~\times$ 2.12 MHz (the 
critical value is $\omega_{crit} = 2\pi~\times$ 2.04 MHz). The width $w$ of the Gaussian mode is equal to the distance
between the central ions, which is 4.1~$\mu$m in the linear array. }
\end{figure}

\begin{figure*}[hbt]
 \includegraphics[width=\textwidth]{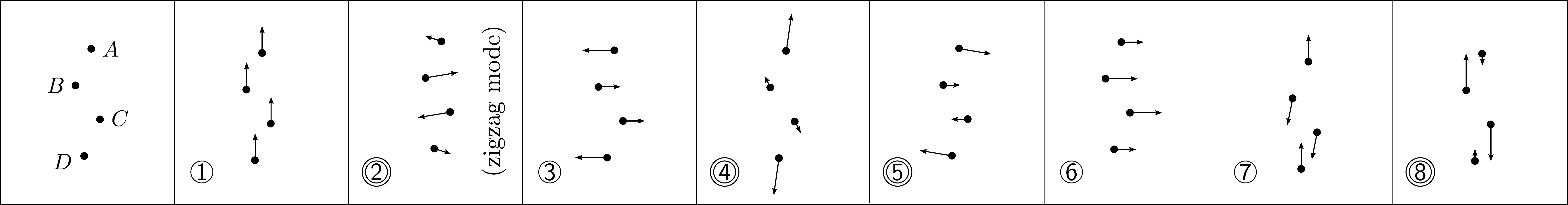} 
 \caption{\label{fig:modes} Schematic representation of the motional modes (ordered by increasing frequency) for the
parameters of Fig. \ref{fig:4ions}(b) with $P=0.074~P_0$ and in the zigzag configuration ($U_0$ is 1/5 of its value for
the linear chain). The positions are not to scale. The modes that couple to the cavity mode are indicated by a double
circle.}
\end{figure*}

The tools developed so far are now employed to characterize the steady state of the ion chain coupled with the cavity
mode. We consider the parameter regime for which multistable solutions are found, corresponding to the choice of
detunings identified in Sec. \ref{sec:bistable}, namely, $\Delta_c=0$, $\Delta_0\gg\gamma$, so that $\Delta_{\rm eff}
<0$. In this regime, photon scattering cools the normal modes coupled to the resonator field, thus providing an instance
of cavity cooling \cite{Vuletic_Chu_PRL_2000}. As we will show, the strong coupling between field and motional
fluctuations leads to distinct signals at the cavity output. A peculiar feature of the stationary state is the
entanglement between light and crystalline vibrations, which is found for large enough cooperativity 
and close to the structural instability. 

The analytical results presented in this Section are valid for a chain of an arbitrary number $N$ of ions. In order to
facilitate the interpretation of the results, the plots will be evaluated for a chain of $N=4$ ions.  Figure
\ref{fig:4ions} displays the equilibrium positions along the $x$ axis for different values of the cooperativity and as a
function of the pump power $P$. In the right panels one clearly observes bistable regions when the cooperativity is
sufficiently large. 

In the following, cooling and entanglement are discussed for two exemplary configurations: in the first configuration,
which we will denote by ``symmetric configuration'', the chain is centered with respect to the cavity axis. The
corresponding equilibrium positions as a function of $P$ are displayed in the upper plots of Fig. \ref{fig:4ions}. In
this case, due to the symmetry under reflection about the center of the chain, some vibrational modes are decoupled from
the cavity mode. These are, for the zigzag chain, the modes labeled by 1, 3, 6, 7 in Fig. \ref{fig:modes}. In the second
case, which we denote by ``asymmetric configuration'', the chain is slightly displaced in $y$ direction. The equilibrium
positions of this configuration as a function of $P$ are shown in the lower plots of Fig. \ref{fig:4ions}.  In this
second case the symmetry  is broken and all motional modes can couple to (and thus be cooled by) the cavity field.

\subsection{Stationary state of the fluctuations} \label{sec:covariance}

We now study the stationary state of the system composed by the cavity fluctuations and the ions' vibrations about their
classical equilibrium values. In general, the knowledge of the density matrix is required. If the initial state is
assumed to be Gaussian, it is Gaussian at all times for the noise and dynamics we consider and is hence fully
characterized by the first and second moments of the observables $Q_a, P_a$ and $Q_n, P_n$ composing the vector $\vec X$
of Eq. (\ref{eq:Xdef}). For input operators and an initial state with zero mean value, the state is completely described
by the covariance matrix \cite{Ferraro_etal_2005}, whose elements in this case read $C_{\alpha\beta} = \langle X_\alpha
X_\beta+X_\beta X_\alpha\rangle$, where $\alpha,\beta$ run over all the components of vector $\vec X$. The covariance
matrix at steady state can be computed using Eq. (\ref{eq:general solution}) in the limit $t\to\infty$.

\begin{figure}[th]
 \includegraphics[width=0.83\columnwidth]{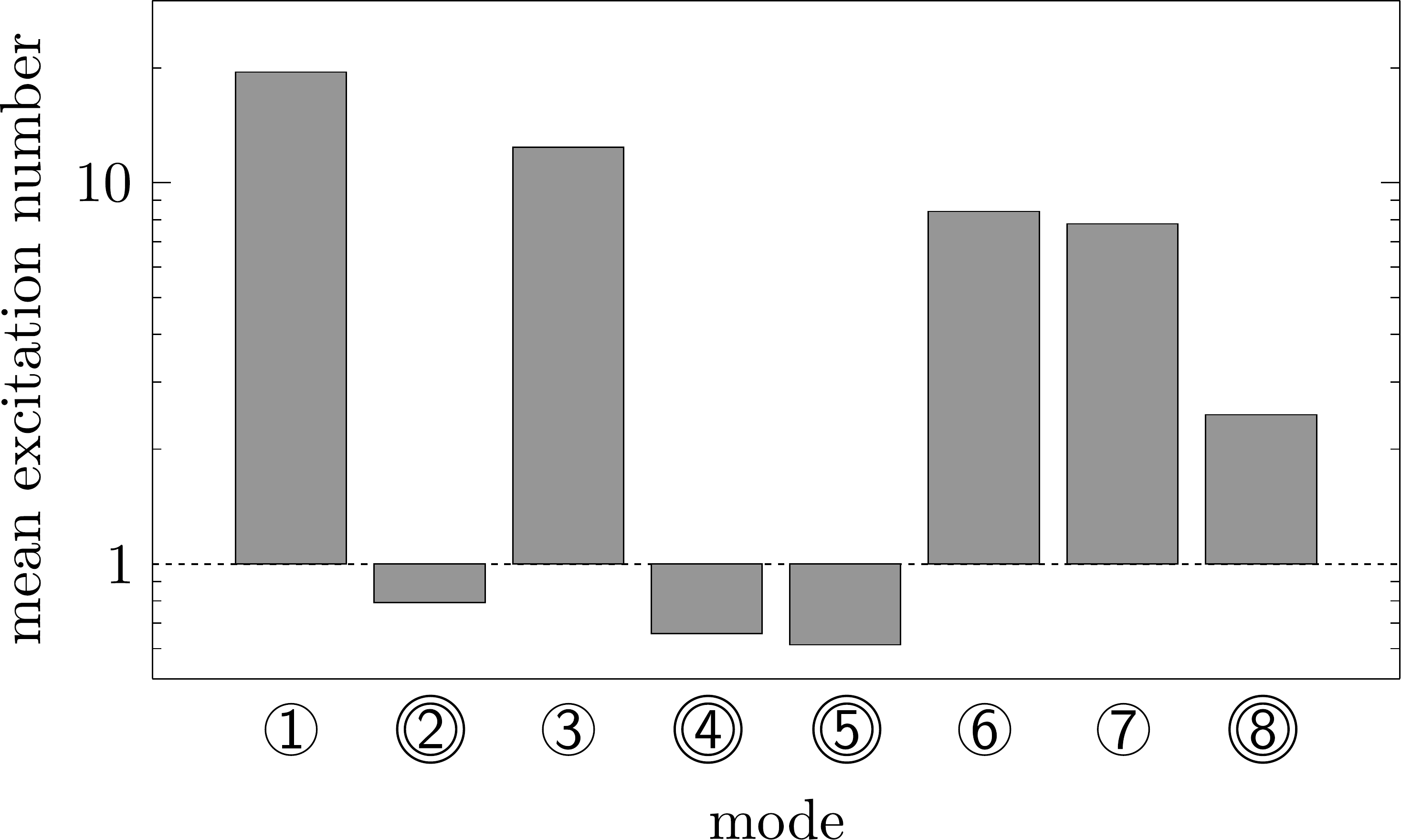} 
 \caption{\label{fig:mode_occupation} Mean excitation number $\langle b^\dagger_n b_n\rangle$ for each of the motional
modes in Fig. \ref{fig:modes}. The dotted line indicates the value $\langle b^\dagger_n b_n\rangle=1$ (the vertical axis
is in logarithmic scale). 
}
\end{figure}

With these results at hand it is straightforward to evaluate the total kinetic energy, mean mode populations, and
squeezing. Figure \ref{fig:mode_occupation} displays the stationary mean occupation number $\langle b^\dagger_n
b_n\rangle$ of the $n$-th motional mode for a zigzag chain in the symmetric configuration for parameters for
which a bistable configuration is found in Fig.~\ref{fig:4ions}(b). Here, it is assumed that the initial state is
thermal with temperature $T=1$ mK. The modes which are cooled by photon scattering are the ones in Fig. \ref{fig:modes}
labeled by 2, 4, 5, 8. At steady state most of them are significantly colder than the uncoupled ones, with mean
excitation number smaller than unity. We note that the mode with the highest frequency is colder than the initial
temperature but still at a significantly higher temperature than the other modes coupling with the cavity fluctuations.
Indeed, this mode corresponds mostly to motion in the $y$ direction, thus weakly coupled to the 
cavity field. 

\subsection{Spectrum of light at the cavity output} \label{sec:spectrum}

\begin{figure}[hbt]
 \includegraphics[width=\columnwidth]{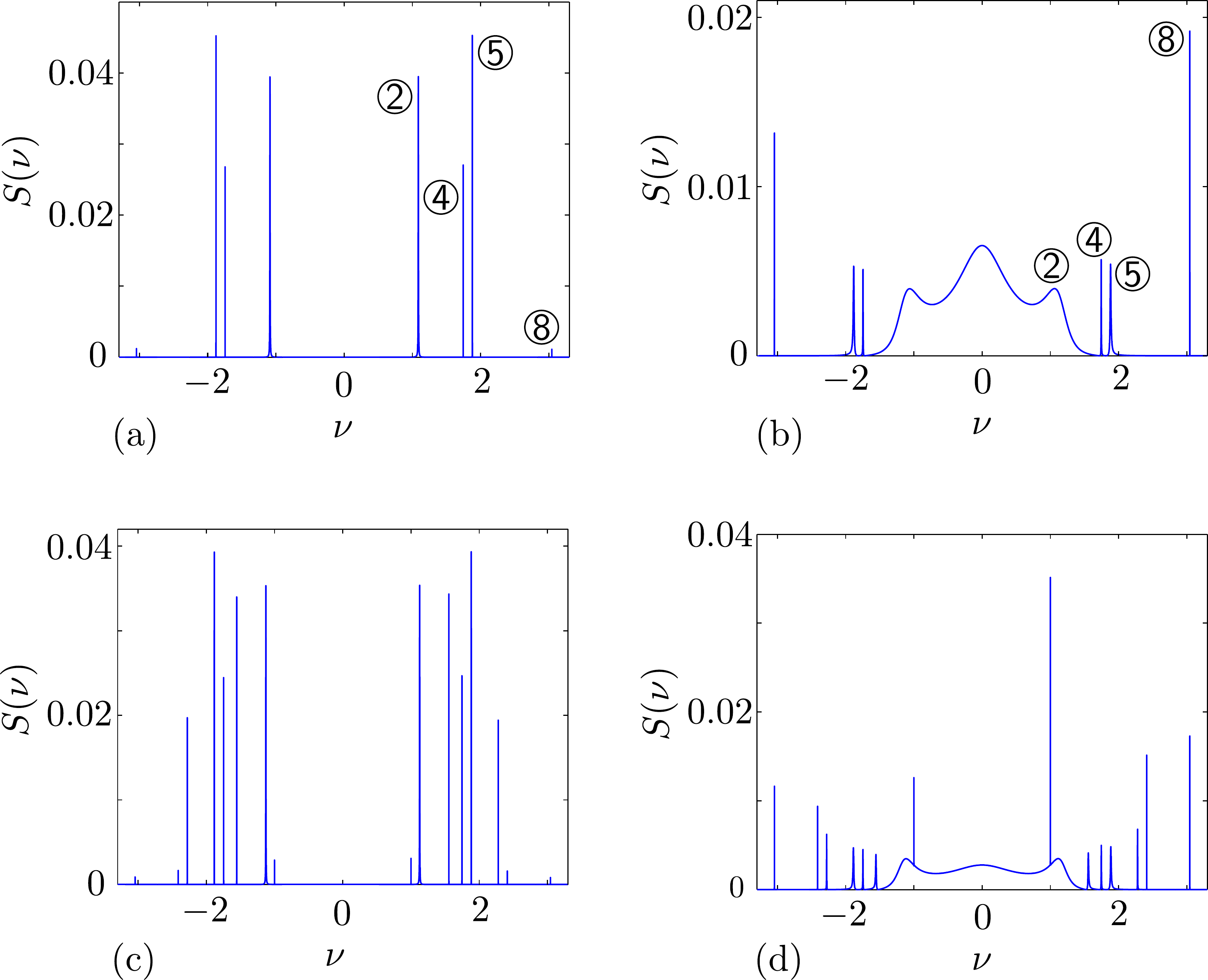}
 \caption{\label{fig:spectra} (color online) Spectrum $S(\nu)$ of the field at the cavity output at steady-state for a
zigzag of four ions (the Rayleigh peak is not shown).  $S(\nu)$ is evaluated from Eq. \eqref{eq:spectrum} and reported
in units of $\omega_y^{-1}$, while $\nu$ is in units of $\omega_y$. The upper (lower) panels refer to the symmetric
(asymmetric) configuration, where in the asymmetric case the chain is displaced with respect to the cavity axis by 0.28
times the distance between the central ions. The parameters are the same as in Fig. \ref{fig:4ions} except for (a)
$\mathcal C=0.3$ and $P=0.54$; (b) $\mathcal C=3$ and $P=0.074$; (c) $\mathcal C=0.3$ and $P=0.62$; (d) $\mathcal C=3$
and $P=0.084$ (the values of $P$ are such that $U_0$ is 1/5 of the value it takes when the ions form a linear array).
The motion is assumed to couple to a thermal bath at temperature 1mK and thermalization rate $\Gamma_n = 100$~s$^{-1}$
for all modes. }
\end{figure}

Information about the dynamical properties of the system in the stationary state can be extracted from the spectrum of
the field at the cavity output. Such measurement has been performed, for instance, in Ref. \cite{Brahms}.  Here, we
determine the spectrum of the field scattered by the ions at the steady state of the cavity cooling
dynamics. We denote by  $S(\nu)\propto  \langle \tilde{a}_{\rm out}(\nu)^\dagger \tilde{a}_{\rm out} (\nu) \rangle$ the
component of the spectrum at frequency $\nu=\omega-\omega_p$, with $\tilde{a}_{\rm out}$ the Fourier transform of the
field at the cavity output, $a_{\rm out} = a_{\rm in} + \sqrt{2\kappa} \, a$. Using that $a=\bar a+\delta a$, the
quantum component of the spectrum reads
\beq
S(\nu) =\frac{\langle \delta \tilde{a} (\nu)^\dagger \delta \tilde{a} (\nu) \rangle}{\bar a^2} \,,
\eeq
where $\delta \tilde{a} (\nu)$ is the Fourier transform of $\delta a$, $\delta \tilde{a} (\nu)^\dagger$ is the Hermitian
conjugate of $\delta \tilde{a} (\nu)$, and we omit the Rayleigh peak at $\nu=0$, i.e., $\omega=\omega_p$, which
corresponds to the classical part \cite{Bienert_etal_PRA_2004}. For the setup we consider, when the ions form a linear
string they are at an antinode of the cavity field and the coupling of the vibrational modes to the cavity is negligible
in the Lamb-Dicke regime (see Sec. \ref{sec:fluctuations}). Therefore, for the linear chain the steady-state spectrum
exhibits only the Rayleigh peak. When the chain forms a zigzag, the spectrum of the field at the cavity output reads 
\begin{multline}
\label{eq:spectrum}
S(\nu)= S_0(\nu)
\Bigg\{ \frac{4\kappa \, |\theta(\nu)|^2 \, \bar a^2}{\kappa^2+(\nu-\Delta_{\rm eff})^2} \\
+ \sum_n c_n^2 \Gamma_n \left[ \frac{N_n}{\Gamma_n^2 + (\omega_n-\nu)^2} + \frac{N_n + 1}{\Gamma_n^2 + (\omega_n+\nu)^2}
\right] \Bigg\}\,,
\end{multline} 
and shows the appearance of Fourier components at the frequency of the vibrational modes coupling with the cavity.
Expression \eqref{eq:spectrum} has been derived by formally integrating Eq. (\ref{eq:b}) for the evolution of $b_n(t)$,
replacing in the evolution of the field fluctuations given by Eq. (\ref{eq:a}), and solving the corresponding system in
Fourier space. An alternative form is provided in Appendix \ref{App:B}. We now discuss the individual terms on the
right-hand side of Eq.~(\ref{eq:spectrum}). The first term inside the curly brackets is the contribution due to the
coupling of the quantum vacuum with the cavity field fluctuations, where 
\beq
\theta(\nu) = \sum_n \frac{c_n^2 \, \omega_n}{\omega_n^2+(\gamma_n-i\nu)^2}\,
\eeq
is  a consequence of the coupling with the vibrational modes. The other terms inside the curly brackets are due to the
thermal noise on the vibrational modes, and their contribution scales with the corresponding strength of the coupling
with the cavity-field fluctuations. The common prefactor reads
\beq
S_0(\nu)=\frac{2}{\kappa^2 + (\nu+\Delta_{\rm eff})^2} \left|1+\frac{4\theta(\nu) \, \Delta_{\rm eff} \, \bar
a^2}{(\kappa-i\nu)^2+\Delta_{\rm eff}^2}\right|^{-2}\,.
\eeq
This function becomes a Lorentz curve when ${\mathcal C}\ll 1$. This functional behaviour is strongly modified when the
cooperativity is increased: Then, motional and quantum noise do not simply add up, but nonlinearly mix to determine the
spectral properties of the output field.  Figures \ref{fig:spectra}(a) and (b) display the spectra for a chain of four
ions in the symmetric configuration, and for two different values of the cooperativity. Note that the Rayleigh peak is
not shown. The spectral lines observed in Fig. \ref{fig:spectra}(a) correspond to the sidebands of the normal modes
which couple to the cavity field fluctuations. As $\mathcal C$ is increased, Fig. \ref{fig:spectra}(b), the spectral
lines change the relative heights, width, and shape. We note the asymmetry in the spectra with respect to $\nu=0$: This
is due to the (weak) coupling of the ions' motion to the thermal bath\footnote{In the limit $\Gamma_n \to 0$ for all
modes, when the only dissipative mechanism in the evolution are the cavity 
losses, 
the spectrum reads:
\beq \nonumber
S(\nu) = \frac{8\kappa \, \bar a^2 |F(\nu)|^2}{\Big|4 \Delta_{\rm eff} \, \bar a^2 F(\nu) + [\Delta_{\rm eff}^2 +
(\kappa-{\rm i}\nu)^2] \prod_{n} (\omega_n^2-\nu^2) \Big|^2},
\eeq
where
\beq \nonumber
F(\nu) = \sum_n c_n^2 \omega_n \prod_{n'\neq n} (\omega_{n'}^2-\nu^2).
\eeq
Hence, the output spectrum is symmetric under $\nu \to -\nu$. In fact, in this case at steady-state there is no net
transfer of energy from the motion to the light field \cite{Wilson-Rae_etal_NJP_2008}.}. 
The broadening at large cooperativity is a consequence of the vacuum input noise and indicates the rate at which the
cavity cools the corresponding vibrational mode \cite{Bienert_etal_PRA_2004}. It is accompanied by the appearance of
Fano-like resonances which result from the dispersive effect of the cavity back-action and are a signature of quantum
interference in the fluctuations of motion and field \cite{Fano_PR_1961}. This interference is due to quantum
correlations established by the dynamics described in Eqs. (\ref{eq:a})-(\ref{eq:b}), and is reminescent of dynamics
studied in optomechanical systems \cite{Pirandola_etal_PRA_2003}. For comparison, Figs. \ref{fig:spectra}(c) and (d)
show the spectrum for the case in the asymmetric configuration: There, all vibrational modes couple with the cavity
field and peaks corresponding to each of the eight motional modes can be identified. 

\subsection{Entanglement} \label{sec:entanglement}

\begin{figure}[ht]
 \includegraphics[width=\columnwidth]{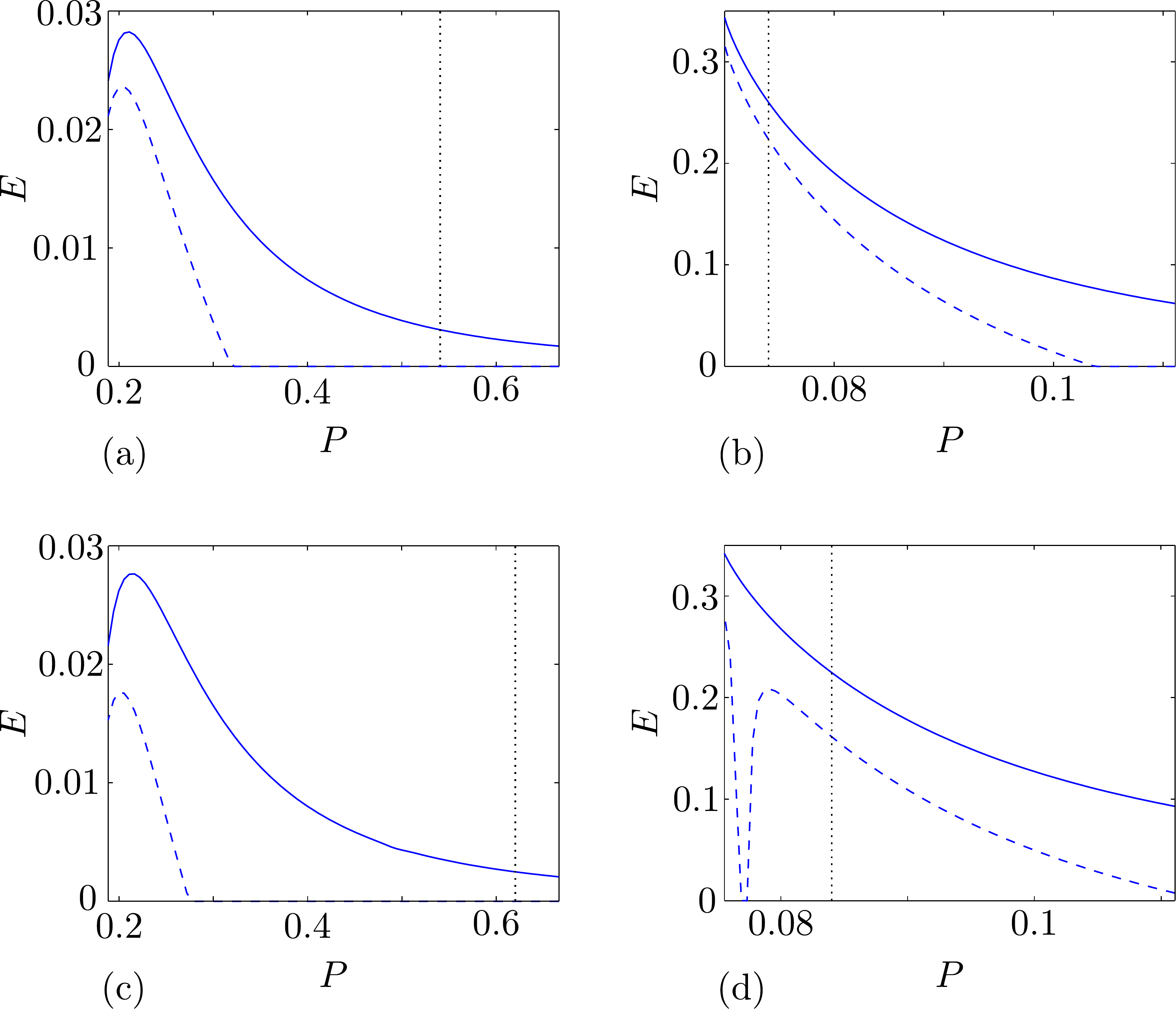}
 \caption{\label{fig:entanglement} (color online) Entanglement between field and motional fluctuations as a function of
pumping power $P$. The entanglement is given by the logarithmic negativity. The continuous line shows the entanglement
between the cavity fluctuations and the whole set of vibrational modes, the dashed line the entanglement between the
cavity and the zigzag mode only. The parameters are the same as in Fig. \ref{fig:4ions} for (a)-(d), respectively, but
the results here refer only to zigzag configurations and within the Lamb-Dicke regime. The dotted vertical lines
indicate the values for which the spectra in Fig.~\ref{fig:spectra} are evaluated.}
\end{figure}

The optomechanical coupling between a cavity mode and an ultracold atomic ensemble has been demonstrated to give rise to
non-classical light \cite{Brooks}. We now show that in our system this coupling generates stationary entanglement
between the field fluctuations and the vibrations of the ions about their equilibrium positions. We analyze the presence
of entanglement in the system across bipartitions, as quantified by the logarithmic negativity
\cite{Vidal_Werner_PRA_2002}. Figure \ref{fig:entanglement} displays the logarithmic negativity for the zigzag chain of
four ions as a function of the pumping power. The plots show the entanglement of the cavity field with the whole set of
motional modes and with the zigzag mode only, and correspond to the cases in Fig.~\ref{fig:4ions} for pumping powers $P$
at which the zigzag chain is a stable configuration. One clearly observes that quantum correlations between field and
vibrations are particularly important for large cooperativity and close to the mechanical 
instability
of the zigzag chain. Besides, correlations are mostly built between the cavity and the zigzag mode, as one can also
infer from  the broadening of the corresponding spectral peak in Figs. \ref{fig:spectra}(b) and (d), indicating the
strong coupling with the cavity field. The results for the asymmetric setup are qualitatively similar to the symmetric
case, except that the entanglement between cavity and zigzag mode is smaller. In particular, in Fig.
\ref{fig:entanglement}(d) one observes that it vanishes for a value of the power that corresponds to an avoided level
crossing between two vibrational modes. The plots on the right pannels correspond to an effective cooperativity which is
ten times as large as in the left pannels. 

A systematic comparison of how entanglement scales with the cooperativity requires to set some criteria. A naive
comparison between the values of entanglement as the cooperativity is increased shows that -for the powers $P$ shown in
Fig. \ref{fig:entanglement}-
entanglement is approximatively an order of magnitude larger when the effective cooperativity is increased by the same
amount. However, such comparison has some drawbacks. For instance, the transverse equilibrium positions of
the ions are different, and for large cooperativity there are no solutions corresponding to zigzags with arbitrarily
small amplitude. In addition, our treatment becomes invalid when the Lamb-Dicke approximation does not apply. A possible
comparison is then to consider the entanglement between light and vibrations when the ions are at the same equilibrium
positions but for different cooperativities. An example is given by the vertical dotted lines in the left and right
plots of Fig. \ref{fig:entanglement}, which indicate the values of $P$ at which the same 
equilibrium structures are found at different values of the cooperativity. In this case, we again verify that
entanglement increases with the cooperativity. In particular, for the same transverse equilibrium displacements,
increasing the cooperativity by a factor of ten can increase the logarithmic negativity by almost two orders of
magnitude. 

\section{Conclusions} \label{sec:conclusions}

An ion chain strongly coupled with the mode of a high-finesse resonator exhibits bistability in the structural
properties. Bistability emerges from the interplay between the Coulomb repulsion and the mechanical effects of the
cavity mode, which mediates a long-range interaction between the ions coupling with it. Bistable configurations, with
coexistence of the linear and the zigzag chain, are found for a range of values of the intensity of the laser driving
the cavity and determining the mean value of intracavity photons, and are associated with histeretical behaviour of the
intensity of the field at the cavity output. We have studied the quantum fluctuations about one of the bistable
configurations, in a setup in which the vibrations of the linear chain weakly couple with the cavity fluctuations while
the zigzag maximizes the mechanical coupling with the field. For the zigzag chain, we have found that the cavity field
cools the vibrational modes coupled with the field fluctuations and becomes entangled with 
them. This entanglement is a stationary property of the system, which is hence cooled into a nonclassical state of light
and vibrations. We identify signatures of this behaviour in the spectral features of the light at the cavity output when
the photons are emitted by the scattering events at the steady state of the cavity cooling dynamics.  

The analysis has been performed in the semiclassical limit, assuming that the crystal ground state is classical and the
mean intracavity photon number is large. Nevertheless, in the strong coupling regime one could observe structural
changes at the single photon level. A further interesting outlook is to explore the dynamics of the coupling between the
cavity photons and the chain close to the linear-zigzag quantum phase transition \cite{Shimshoni}. Here, novel quantum
states of matter and light are expected with properties yet to be determined. 

\acknowledgements

The authors thank I. Leroux, F. Cartarius, M. Bie\-nert, S. Fishman, E. Kajari, and O. Mishina, for fruitful
discussions. This work was partially supported by by the European Commission (STREP PICC, COST action IOTA, integrating
project AQUTE), the BMBF project QuORep, the Alexander-von-Humboldt and the German Research Foundations.

\appendix

\section{Stability conditions for the fluctuations} \label{ap:stability}

In this Appendix we analyze the stability of the linear system (\ref{eq:linear compact}) governing the evolution of the
fluctuations. Stability is guaranteed by the negativity of the real parts of all eigenvalues of the matrix $M$ defined
in Eq. (\ref{eq:M}). An equation for the eigenvalues can be derived generalizing the procedure in
\cite{Ullersma_Physica_1966}; the assumption on how the input noise acts on the motional modes in (\ref{eq:b}) is not
crucial for the derivation. We look for solutions of $M \vec X = \lambda \vec X$, which implies:
\begin{align}
\label{eq:cond1}
&A_n X_a + M_n X_n = \lambda X_n \quad \forall n \,,\\
\label{eq:cond2}
&M_a X_a + \sum_n A_n X_n = \lambda X_a \,.
\end{align}
The matrices $M_n$ are diagonal and have eigenvalues $-\Gamma_n \pm {\rm i} \omega_n$. If $\lambda = -\Gamma_n \pm {\rm
i} \omega_n$, then Eq. (\ref{eq:cond1}) implies that either $c_n=0$ or $X_a=0$. In the first case, the mode is decoupled
from the evolution of the rest of the system; such modes will be ignored in the following. The second case can only
arise if all the modes that couple to the cavity have the same eigenvalues $-\Gamma_n \pm {\rm i} \omega_n$; in this
case, it is possible to find linear combinations of these modes that get decoupled. This can be solved by redefining the
modes, and from now on we only deal with modes for which $c_n$ does not vanish and which have non-degenerate eigenvalues
$-\Gamma_n \pm {\rm i} \omega_n$. Then, the eigenvalues $\lambda$ of $M$ cannot equal $-\Gamma_n \pm {\rm i} \omega_n$,
and the matrix $\lambda \mathbb{I} - M_n$ can be inverted so that Eq. (\ref{eq:cond1}) gives:
\beq
X_n = (\lambda \mathbb{I} - M_n)^{-1} X_a \quad \forall n \,.
\eeq
Plugging this in Eq. (\ref{eq:cond2}), one obtains the generalized eigenvalue equation:
\beq
\left[ M_a +\sum_n A_n (\lambda \mathbb{I} - M_n)^{-1} A_n \right] X_a = \lambda X_a \,.
\eeq

In order to find non-zero solutions for this equation (which means modes in which the cavity degrees of freedom are
involved) the eigenvalues $\lambda$ have to satisfy:
\beq
\det\left[ M_a -\lambda \mathbb{I} +\sum_n A_n (\lambda \mathbb{I} - M_n)^{-1} A_n \right] = 0 \,,
\eeq
leading to the condition Eq. (\ref{eq:eigenvalues}).
Decomposing the eigenvalues in real and imaginary parts, $\lambda = \alpha + {\rm i} \beta$, Eq.~(\ref{eq:eigenvalues})
can be split in its real and imaginary parts:
\begin{widetext}
\begin{align}
\label{eq:eig real}
\Delta_{\rm eff}^2 + (\kappa+\alpha)^2 - \beta^2 &= -4 \Delta_{\rm eff} \bar a^2 \sum_n \frac{c_n^2 \omega_n ~
[\omega_n^2 + (\alpha+\Gamma_n)^2-\beta^2]}{[\omega_n^2 + (\alpha+\Gamma_n)^2-\beta^2]^2 +4\beta^2(\alpha+\Gamma_n)^2}
\,,\\
\label{eq:eig imag}
(\kappa+\alpha)\beta &= 4 \Delta_{\rm eff} \bar a^2 \beta \sum_n \frac{c_n^2 \omega_n ~ (\alpha+\Gamma_n)}{[\omega_n^2 +
(\alpha+\Gamma_n)^2-\beta^2]^2 +4\beta^2(\alpha+\Gamma_n)^2} \,.
\end{align}
\end{widetext}
We first consider the real eigenvalues, with $\beta=0$. Since $c_n$ is real and $\omega_n$, $\Gamma_n$ are non-negative,
Eq. (\ref{eq:eig real}) implies that for $\Delta_{\rm eff}>0$, when the pump is above the effective resonance of the
cavity, there can be no such solutions. On the other hand, when $\Delta_{\rm eff}<0$, stability conditions can be found
for the solutions observing that in the region $\alpha>0$, the left-hand-side in Eq. (\ref{eq:eig real}) is increasing,
going to infinity as $\alpha\to\infty$, while the right-hand-side is decreasing, with zero as limit. This leads to the
following necessary condition for stability:
\beq \label{eq:stabilityap}
\Delta_{\rm eff}^2+ \kappa^2 \geq -4\Delta_{\rm eff} \, \bar a^2 \sum_n \frac{c_n^2 \omega_n}{\omega_n^2 + \Gamma_n^2}
\quad (\Delta_{\rm eff}<0) \,,
\eeq
which reduces to the one found in \cite{Ullersma_Physica_1966} when $\kappa=\Gamma_n=0$. If this condition is not
satisfied, there is exactly one real positive eigenvalue, corresponding to an unstable mode.

We now turn to the complex eigenvalues, for which $\beta\neq0$. 
For $\Delta_{\rm eff}<0$, the two sides of Eq. (\ref{eq:eig imag}) can only have the same sign if the following
condition is satisfied:
\beq
\alpha \in (\min\{-\kappa, -\Gamma_n\}, \max\{-\kappa, -\Gamma_n\}) \quad (\Delta_{\rm eff}<0)
\eeq
(we are actually interested in the case $\kappa>\Gamma_n$, so that the lower limit is $-\kappa$). This means that the
damping rates for each of the collective modes are within the interval determined by the loss rates of the modes
composing the system (this holds only for the complex eigenvalues; the purely real ones need not obey this condition).
It also proves that if $\Delta_{\rm eff}<0$ the condition (\ref{eq:stabilityap}) is necessary and sufficient to have no
eigenvalues with positive real parts. For \mbox{$\Delta_{\rm eff}>0$}, one finds instead:
\beq
\alpha \in \mathbb{R} \setminus [\min\{-\kappa, -\Gamma_n\}, \max\{-\kappa, -\Gamma_n\}] \quad (\Delta_{\rm eff}>0) \,.
\eeq
Thus, if the only environment corresponds to cavity losses (i.e. $\Gamma_n = 0 ~ \forall \, n$), the real parts of the
eigenvalues must be either below $-\kappa$ or above 0. Taking into account that in this case the trace of $M$ is
$-2\kappa$, for every pair of complex eigenvalues with real part smaller than $-\kappa$ there must also be eigenvalues
with real parts larger than zero, namely unstable solutions. Since for $\Delta_{\rm eff}>0$ there are no purely real
eigenvalues, this indicates that the cavity provides a heating mechanism for at least one mode (but this cavity heating
can be compensated by a cooling environment for the motion to keep the system stable).

\section{Alternative form of the spectrum at the cavity output}
\label{App:B}

As an alternative to the procedure in Section \ref{sec:spectrum}, the steady-state spectrum can be calculated in terms
of the matrices $D$ and $T$ introduced in Section \ref{sec:fluct evolution} for the diagonalization of the problem. This
is done by Fourier-transforming the solution (\ref{eq:general solution}) for the evolution of the cavity fluctuations in
terms of the collective eigenvectors with $t_0\to-\infty$, and the resulting expression is:
\begin{multline} \label{eq:spectrum_alternative}
S(\nu) = \frac{\bar a^2}{2\pi} \Bigg\{ 2\kappa \left| \left(\tilde T \frac{1}{D+i\nu} \tilde T^{-1} \right)_{\!\delta a,
\delta a^\dagger} \, \right|^2 \\
+  \sum_n 2\Gamma_n (\bar N_n +1) \left| \left(\tilde T \frac{1}{D+i\nu} \tilde T^{-1} \right)_{\!\delta a, b_n^\dagger}
\, \right|^2 \\
+  \sum_n 2\Gamma_n \bar N_n \left| \left(\tilde T \frac{1}{D+i\nu} \tilde T^{-1} \right)_{\!\delta a, b_n} \, \right|^2
\Bigg\}
\end{multline}
where $\tilde T$ is the matrix performing the transformation from the creation and annihilation operators $\{\delta a,
\delta a^\dagger, b_1, b_1^\dagger, \ldots, b_N, b_N^\dagger\}$ into the basis of collective eigenvectors; $\tilde T$
can be obtained from $T$ in a straightforward way using the definitions (\ref{eq:motion quad}) and (\ref{eq:field
quad}). It is apparent in (\ref{eq:spectrum_alternative}) that the eigenvalues of the collective evolution correspond to
poles of $S$ in the complex plane. However, this expression only looks simpler than (\ref{eq:spectrum}) at the expense
of hiding the complexity in the matrices $D, ~\tilde T$.


\end{document}